\documentstyle{article}

\begin{document}

\begin{center}
{\huge\bf On the fundamental structure of quantum mechanics}
\end{center}

\vspace{1cm}
\begin{center}
{\large\bf
F.GHABOUSSI}\\
\end{center}

\begin{center}
\begin{minipage}{8cm}
Department of Physics, University of Konstanz\\
P.O. Box 5560, D 78434 Konstanz, Germany\\
E-mail: ghabousi@kaluza.physik.uni-konstanz.de
\end{minipage}
\end{center}

\vspace{1cm}

\begin{center}
{\large{\bf Abstract}}
\end{center}
We show that there is no real difference between mathematical models of quantum mechanics and classical mechanics concerning integrable dynamical systems because the main difference between them results from their different interpretations.
\begin{center}
\begin{minipage}{12cm}

\end{minipage}
\end{center}

\newpage
Although it is customary to believe that there are main differences even between mathematical models of quantum mechanics (QM) and classical mechanics (CM) but a careful comparison of mathematical structures of both models shows that they are mathematically equal and only their classical physical and quantum physical interpretations are different. Note in this respect that just QM is mainly a mathematical model defined on Hilbert function space of square integrable/square summable functions. Thus as we discuss in the following even the Fourier theory of square integrable Hilbert space functions appears also in CM.

In order to show this equivalence let us recapitulate the mathematical structures of the main four differences between QM and CM. Namely the presence of commutators among operators, uncertainty relations, interference for microscopic particles and entanglement in QM and their absence in CM.

From logical point of view these differences can be summarized in one difference with different representations in view of the intrinsic relations among the commutators, uncertainty relations, interference and entanglement in QM. Thus in view of basic Hilbert space equivalence between Heisenberg's operator formalism and Schroedinger's wave formalism \cite{vn} not only the commutator relations and uncertainty relations are connected with each other but also the theoretical descriptions of interference and entanglement are related with the uncertainty principle \cite{quel}. Insofar proving the existence of commutator relations in CM would be enough to show the existence of connected uncertainty relations and thereby the existence of rest two mentioned  QM properties in CM. Nevertheless we will show in the following that beside that the structural stability theory of CM includes similar relations as the uncertainty relation of QM.

To prove this note that although usually one calculates Poisson brackets in CM with positions and momentum variables as the phase space variables but this approach is equivalent to calculation with vector fields and one may describe CM by poisson brackets of hamiltonian vector fields as in the symplectic geometry \cite{arnold}. Thus there are commutator relations among vector fields in CM $[L_B, L_A] = [B_i \displaystyle{\frac{\partial}{\partial_i}}, A_j \displaystyle{\frac{\partial}{\partial_j}}]; \ i, j, k = 1, 2, 3$ \cite{arnold} which play the same role as the commutators among hamiltonian vector field operators in QM. Insofar the constant QM commutators among position and momentum vector field operators in momentum or position representation, e. g. $[\hat{Q_i} , \hat{P_j} ] = i \hbar \delta_{ij}$ are up to constant factors equal to the vector field descriptions of Poisson brackets of vector fields $ [L_{Q_i}, L_{P_j}] = \delta_{ij}$ as in the symplectic CM \cite{am}, \cite{arnold}. In other words specially the commutator algebras $[L_A, L_B] = L_C, L_C := cte.$ are given for vector fields both in CM and QM. Accordingly one may expect that in view of the presence of commutators in CM and the well known connection between uncertainty relations and commutators in QM also uncertainty relations should be present in CM. Indeed this is the case and we will show it in the context of structural stability theory of CM \cite{am} in the following. Note for technical implementation of uncertainty relations in QM from commutators with the help of the Hilbert space functions that the Fourier theory of square integrable functions are used also in CM specially to prove the structural stability or integrability of CM systems by Hamiltonian perturbation theory \cite{am}, \cite{al}. Insofar even the QM typical Hilbert space structure appears also in CM as one may expected for any integrable system.

To find the existence of uncertainty relation in CM we conjecture that one may consider the uncertainties for canonical variables as small perturbations in the sense of theory of structural stability \cite{am}. Thus a structurally stable two dimensional dynamical system, i. e. a structurally stable dynamical system with two degrees of freedom is stable against small perturbations of its Hamiltonian $\delta H (p, q)$ as the function of small perturbations of phase space variables $\delta q$ and $\delta p$ \cite{am}. Interpreting $\delta q$ and $\delta p$ as uncertainties for position and momentum variables with a suitable small constant value for their product $\delta q. \delta p := \varepsilon$ one obtains a similar relation in CM as in QM uncertainty relation which differ only by a small constant number $\hbar$. Thus in CM the well known small perturbation of Hamiltonian of a dynamical system can be considered to be caused by small perturbations of its position and momentum content. Obviously for a structurally stable or integrable system, i. e. with an action integral $S_{CM} = \int [p(t) dq (t) - H (p(t), q(t))]dt$ the small perturbations of system are caused by $\delta p$ and $\delta q$. If the system is integrable or structurally stable this means that the perturbed action is $\tilde{S}_{CM} = \int [p(t) dq (t) - H (p(t), q(t))]dt + m. \varepsilon, \ m (cte.) <$ and the equations of motions remain the same describing the same dynamics. The situation is well comparable with QM case where $\Delta p . \Delta q \approx \hbar$ and hence the original Bohr-Sommerfeld quantization of action $S_{QM} = \int [p(t) dq (t) - H (p(t), q(t))]dt = N \hbar$ becomes in view of uncertainties $S_{QM} = \int [p(t) dq (t) - H (p(t), q(t))]dt = (N+m) \hbar$ describing the same dynamics as before. The simple reason for same dynamics, i. e. same dynamical equations against small perturbations in both CM and QM cases is that the variation of small perturbations is vanishing.

Moreover the equality of the mathematical structure of both theories up to constants becomes more obvious if one recalls that Heisenberg used in his invention of QM the same radiation formula attached to electrons, i. e. $\int_\alpha \ or \ \sum_\alpha U_\alpha (n) e^{i \omega (n, n- \alpha) t}$ as the classical radiation $\int_\alpha \ or \ \sum_\alpha U_\alpha (n) e^{i \omega (n) . \alpha t}$ \cite{heis}. The only difference is that in QM case the frequency is a function of difference $\omega(n, n- \alpha)$ whereas in CM case it is the function $\omega(n)$. Considering all this equalities among the mathematical structures of both theories one may conclude that QM is the integrable version of a CM where the integrability constant is assumed to be proportional to the Planck's constant. Also the above mentioned similarity of radiation formulas in CM and QM recalls the above mentioned presence of Fourier theory and square integrable functions in both cases and thereby the equivalence of mathematical structure in CM and QM.

In the same manner also it becomes obvious in the last decades that the interference pattern is no typical property of microscopic QM particles like electrons but it appears also for macroscopic large molecules in CM \cite{me}.

Accordingly in view of theoretical inclusion of uncertainty relations in descriptions of QM interference and entanglement and the existence of interference for macroscopic CM objects one may expects that entanglement can be considered also as a CM property. Thus any description of entanglement in QM is given by linear combinations of trigonometric functions which are also the standard elements in the description of periodic motions in CM. Insofar any theoretical description of entanglement with the help of very classical trigonometric functions may be considered also as a CM descriptions. The only difference here is also the different QM and CM interpretations of the mathematical description of dependency of a two-element system or also more-element systems. Thus also the dynamics of binary star in CM may be considered as entangled and in the same manner also the dynamical states of Earth and Sun in the Earth-Sun system is classically entangled. Then one can not consider the classical state of a subsystem, e. g. Earth alone in CM as a well defined state because Earth as a dynamical subsystem depends on its gravitational interaction at least with the Sun where the influence of rest planets in solar system is considered as a small perturbation against which the Earth-Sun system is structurally stable \cite{am}, \cite{arnold}. In the same manner Sun can not be considered alone as a well defined subsystem in view of its gravitational interactions with Earth and other planets which should be included in a well defined description of dynamics of the Sun. Thus the Earth-Sun as a structurally stable dynamical two-element compound system can be considered as entangled globally similarly as a two-element compound system in QM.
The classical interpretation of entanglement can be given in accordance with the structural stability of dynamical systems with two degrees of freedom \cite{am} in the following way. The transversality condition in structural stable dynamical systems means that in a structurally stable dynamical system with two degrees of freedom the two components of the vector field are transversal \cite{am}. A property which is comparable to the transversality or orthogonality of cosine and sine functions in the QM entanglement models $\oint cos \alpha . sin \alpha d \alpha = 0$ where the QM entanglement is due to the square summability of these functions $cos ^2 \alpha + sin^2 \alpha = 1$ implying their quadratic dependency on each other. Thus in the two dimensional (2D) case the two components of potential differential one form $A = A_m dx^m, \ m = 1, 2$ as Hodge dual to the mentioned vector field may be given as $A_x = K . y, \ A_y = \pm K. x$ where $+ K$ or $- K$ is the harmonic- or constant component of field strength or curvature two form $F_{[mn]} \in \omega^2 (2D):= d A \oplus Harm^2$, respectively \cite{nak}. In polar coordinates these components are given by $A_x =  K . r sin, \ A_y = \pm K . r cos$. Then such a transversal property is inherent for the structural stability of any dynamical system against small perturbations \cite{am}. Insofar we observe here also the relation of discussed classical entanglement to small perturbations as the above discussed classical uncertainties. Thus the constant commutator of discussed transversal components of vector field of mentioned dynamical system is the one we compared above with the constant commutators of QM.

Footnotes and references

\end{document}